\documentclass[twocolumn,showpacs,superscriptaddress,preprintnumbers,amsmath,amssymb,prb,floatfix]{revtex4}

\usepackage{graphicx}
\usepackage{dcolumn}
\usepackage{bm}
\usepackage{amsmath}
\usepackage{amssymb}
\usepackage{booktabs}
\usepackage{color}

\begin{document}

\preprint{APS/123-QED}

\title{Quantum effects on the loss function of Pb(111) thin films: an \textit{ab initio} study.}

\author{X. Zubizarreta}
   \affiliation{Donostia International Physics Center (DIPC), Paseo de Manuel Lardizabal 4, 20018
   San Sebasti\'an/Donostia, Basque Country, Spain}
   \affiliation{Departamento de F\'{\i}sica de Materiales, Facultad de Ciencias Qu\'{\i}micas,
   Universidad del Pa\'{\i}s Vasco/Euskal Herriko Unibertsitatea, Apdo. 1072, 20080 San Sebasti\'an/Donostia,
   Basque Country, Spain}
\author{E. V. Chulkov}
   \affiliation{Donostia International Physics Center (DIPC), Paseo de Manuel Lardizabal 4, 20018
   San Sebasti\'an/Donostia, Basque Country, Spain}
   \affiliation{Departamento de F\'{\i}sica de Materiales, Facultad de Ciencias Qu\'{\i}micas,
   Universidad del Pa\'{\i}s Vasco/Euskal Herriko Unibertsitatea, Apdo. 1072, 20080 San Sebasti\'an/Donostia,
   Basque Country, Spain}
   \affiliation{Centro de F\'{\i}sica de Materiales CFM - Materials Physics Center MPC, Centro Mixto CSIC-UPV/EHU,
   Paseo de Manuel Lardizabal 5, 20018 San Sebasti\'an/Donostia, Basque Country, Spain}
\author{V. M. Silkin}
   \affiliation{Donostia International Physics Center (DIPC), Paseo de Manuel Lardizabal 4, 20018
   San Sebasti\'an/Donostia, Basque Country, Spain}
   \affiliation{Departamento de F\'{\i}sica de Materiales, Facultad de Ciencias Qu\'{\i}micas,
   Universidad del Pa\'{\i}s Vasco/Euskal Herriko Unibertsitatea, Apdo. 1072, 20080 San Sebasti\'an/Donostia, Basque Country,
Spain}
   \affiliation{IKERBASQUE, Basque Foundation for Science, 48011 Bilbao, Spain}
\date{\today}

\begin{abstract}
A theoretical study of the surface energy-loss function of freestanding Pb(111) thin films is presented, starting from the single monolayer case. The calculations are carried applying the linear response theory, with inclusion of the electron band structure by means of a first-principles pseudopotential approach using a supercell scheme. Quantum-size effects on the plasmon modes of the thinnest films are found in qualitative agreement with previous work based on the jellium model. For thicker films, results show a dispersionless mode at all thicknesses, in agreeement with electron energy-loss measurements. For sizeable values of the momentum, the raising of the surface plasmon with increasing thickness is retrieved.
\end{abstract}

\pacs{71.20.Be,71.45.Gm}

\maketitle

\section{INTRODUCTION}

In thin metallic films, the confinement in the direction perpendicular to the film plane gives rise to the quantization of the electronic wave functions. As a result of the appearance of the so-called quantum-well states (QWS),\cite{chiassp00} the properties of the metallic slabs might strongly depend on the exact thickness of the film. This dependence is a purely quantum phenomena known as quantum-size effect (QSE), which often appears as an oscillatory dependence of several physical properties on the film thickness.

Thin lead films exhibit important quantum-size oscillations in the layer-by-layer growth,\cite{hincepl89} first observed by He-atom scattering and attributed to interference with the quantum-well states. The latter modulate the electron density of states (DOS) at the Fermi level $(E_F)$, causing oscillations with varying thickness in the superconducting critical temperature and the upper critical field,\cite{ozernph06,sklyprl11} interlayer distances,\cite{jiaprb06} island height distributions,\cite{oterprb02} zone-center phonon frequencies,\cite{ynduprl08,brauprb09} electronic transport,\cite{jaloprl96} photoemission properties,\cite{kircprb07} work functions \cite{kimpnas10} and quasiparticle lifetimes.\cite{hobrprb09,kirenp10} Also, recently superconductivity was discovered in a single lead monolayer (ML) on silicon.\cite{zhannph10} Thus, lead films have become an important model system for exploring electronic and structural properties of metals on the nanoscale.\cite{jiajpsj07}

However, to the best of our knowledge, there are few experimental studies on the surface dielectric response of Pb thin slabs,\cite{jaloprb02,jalocoma04,puccprb06} and no theoretical works. Thus, the aim of the present work is performing a computational systematic study of the surface energy-loss function of Pb(111) films with varying thickness, starting from the single monolayer case, up to the 15 ML thick slab.

An approximate description of thin film plasmons is given by the solution of the Maxwell equations for the proper geometry.\cite{ritcpr57} It leads to the coupling between the classical surface plasmons of the two different surfaces of the film. The resulting coupled modes of the film disperse as \cite{ritcpr57,yuanprb06,pitarpp07}
\begin{equation}
\omega_{\pm} = \frac{\omega_{p}}{\sqrt{2}}(1 \pm e^{-qL})^{1/2},\label{classical_dispersion}
\end{equation}
where $\omega_{p}$ is the bulk plasmon frequency, which is given by $\omega_p = \sqrt{4\pi n/m^*}$ with $n$ being the average electron density and $m^*$ the electron effective mass, which in terms of the density parameter $r_s$ standing for the average inter-electron distance reads as $\omega_p = \sqrt{3/r_s^3\,m^*}$. The energy splitting between the modes depends on the film thickness L and the in-plane momentum transfer $q$. The low-energy mode $\omega_{-}$ corresponds to a symmetric induced charge profile in the direction perpendicular to the film plane, whereas the high-energy mode $\omega_{+}$ corresponds to an antisymmetric one.\cite{yuanprb06} As L increases, the coupling between the two modes decreases. In the limit L$\gg$1/$q$ the two film modes are decoupled and the two classical surface plasmons of frequency $\omega_{p}/\sqrt{2}$ are retrieved. This model ignores the electronic structure of the film. This is a serious drawback since the ground state electronic structure has been shown to strongly affect the surface response to external perturbations. More detailed classical models showed the dependence of the surface plasmon dispersion on the microscopic details of the surface electronic density profile.\cite{bennprb70,schwprb82} 

On a more quantitative level, the jellium model \cite{langprb70} has been used to study the quantum-mechanical electrodynamical response of metal slabs,\cite{eguiprl83,dobsprb92,schaprb94} gaining basic insight into the nature of electronic excitations of metallic films. As an example, Yuan and Gao have shown,\cite{yuanprb06} using the jellium model with the electron density corresponding to Ag, the disappearance of the antisymmetric mode $\omega_{+}$ for $q \rightarrow 0$ when the film thickness is comparable to the Fermi wavelength. Instead, a few discrete interband peaks were found.\cite{yuanprb06}

A more precise description of the electron band structure in the direction perpendicular to the film plane,\cite{chulss97,chulss99} allowing to describe the surface states which are missing in a jellium model, was recently used to study new collective electronic excitations at metal surfaces \cite{silkprb05,pohlepl10,krasprb10} and thin metal films.\cite{silkprb11} However, the recipe of the improved one-dimensional potential \cite{chulss97,chulss99} can not give a satisfactory description of the electronic structure of Pb(111) films. Thus, in the present work a first-principles approach is used to study the dielectric response of Pb(111) films. Indeed, using an \textit{ab initio} calculation scheme possible anisotropy effects can be studied, which are missing in jellium models or in using the potentials of Refs.~\onlinecite{chulss97} and \onlinecite{chulss99}, as they assume in-plane free-electron-like behavior.

In general, the plasmon modes of the Pb(111) films are found to follow qualitatively the classical dispersion relation Eq.~(\ref{classical_dispersion}). However, for the thinnest slabs QSE are found. A dispersionless mode is found at all thicknesses, replacing in practice the surface plasmon as the short wavelength limit of the symmetric plasmon mode. The raising of the surface plasmon with increasing thickness is found at short wavelengths. Also, comparison of the surface energy-loss function with the momentum transfer \textbf{q} along different high-symmetry directions showed no sizeable anisotropy effects.

The rest of the paper is organized as follows. In Sec. II the details of the \textit{ab initio} calculation of the surface loss function using a supercell scheme are shown. In Sec. III the calculated ground state electronic structure properties are presented, while in Sec. IV the results on the surface loss function are analyzed in detail. Finally, the main conclusions are drawn in Sec. V. Unless otherwise stated, atomic units are used throughout, i.e., $e^2=\hslash=m_e=1$.

\section{CALCULATION METHOD}

When a perturbing electric charge is located far from one side of the film the differential cross section for its scattering with energy $\omega$ and in-plane momentum transfer \textbf{q} is proportional to the imaginary part of the surface response function \textsl{g}$(\textbf{q},\omega$) defined as \cite{persprb85}
\begin{equation}
\textsl{g}(\textbf{q},\omega) = - \dfrac{2\pi}{q}\int dz \int dz' \chi(z,z',\textbf{q},\omega)e^{q(z+z')},\label{g}
\end{equation}
which depends on the film electronic properties only ($q = |\textbf{q}|$). This quantity is relevant in the description of surface collective excitations measured in electron energy-loss experiments.\cite{liebphs87,liebsch97} Here $\chi(z,z',\textbf{q},\omega)$ is the density response function of an interacting electron system that determines, within linear response theory, the electron density $n^{\rm{ind}}(z,\textbf{q},\omega)$ induced in the system by an external potential $V^{\rm{ext}}(z,\textbf{q},\omega)$ according to
\begin{equation}
n^{\rm{ind}}(z,\textbf{q},\omega) = \int dz'\chi(z,z',\textbf{q},\omega)V^{\rm{ext}}(z',\textbf{q},\omega).\label{nind}
\end{equation}
The collective electronic excitations in thin films then can be traced to the peaks in the surface loss function defined as the imaginary part of \textsl{g}, Im[\textsl{g}($\textbf{q},\omega$)].

\begin{figure*}[t]
\includegraphics[width=1.0\textwidth]{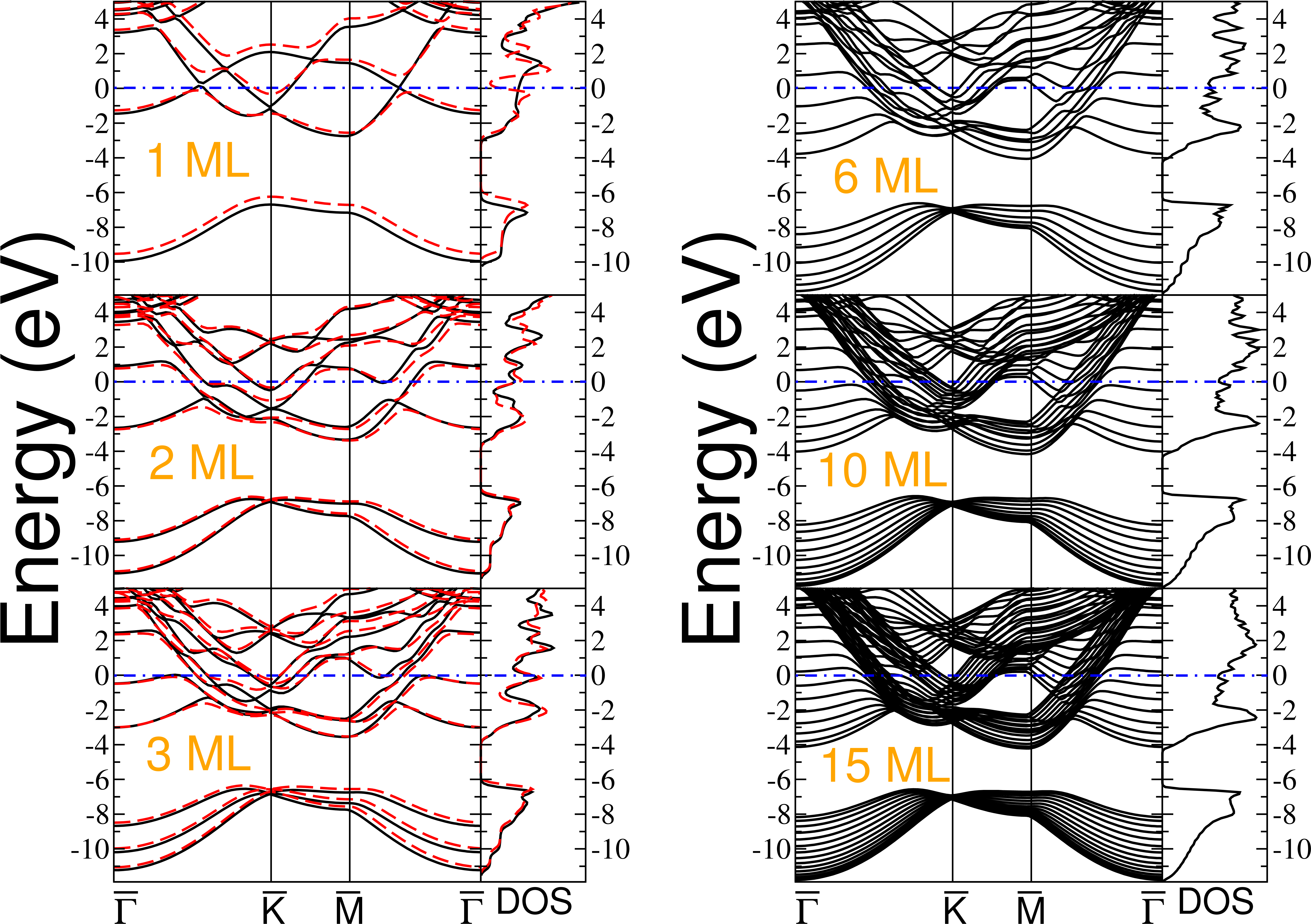}
\caption{(Color online) Calculated band structure  and density of states of Pb(111) freestanding films 1 - 3, 6, 10 and 15 ML thick, with (solid lines) and without (dashed lines) spin-orbit (SO) coupling in the Hamiltonian. The horizontal dashed-dotted lines represent the Fermi level of each film. DOS in arbitrary units.}\label{bs}
\end{figure*}

In the framework of time-dependent density functional theory,\cite{rungeprl84,petersprl96} $\chi$ is the solution of the integral equation
\begin{multline}
\chi(z,z',\textbf{q},\omega) = \chi^{0}(z,z',\textbf{q},\omega) + \int dz_{1} \int dz_{2}\chi^{0}(z,z',\textbf{q},\omega)
\\ \times [v_{c}(z_{1},z_{2},\textbf{q})+ K_{XC}(z_{1},z_{2},\textbf{q},\omega)]\chi(z_{2},z',\textbf{q},\omega),\label{chi_int} 
\end{multline}
with $\chi^{0}$ being the response function of the noninteracting Kohn-Sham electrons. In Eq.~(\ref{nind}) $v_{c}(z,z',\textbf{q}) = -\frac{2\pi}{q}e^{q|z-z'|}$ stands for the two-dimensional (2D) Fourier transform of the bare Coulomb potential and $K_{XC}$ accounts for the exchange-correlation (XC) effects. In the present work, we use the random-phase approximation (RPA) where $K_{XC}$ is set to zero, i.e., the dynamical short-range exchange-correlation effects are ignored. Previous studies of collective excitations at the surfaces \cite{nagao10,nagaprl01,tsuess91,silkprl04} and in the bulk \cite{aryaprl94,krasprb99} of many “metallic” systems suggest that XC effects should have little impact on the study of Pb films. 

For a periodic system, the polarizability can be expressed as a matrix in the basis of the reciprocal space vectors $\left\lbrace\textbf{G}\right\rbrace$. As a consequence, Eq.~(\ref{chi_int}) becomes a matrix equation. Then, once the ground state has been obtained, the starting point of the calculation of the surface response function is the evaluation of the matrix elements of the noninteracting polarizability
\begin{widetext}
\begin{equation}
\chi^{0}_{\textbf{G},\textbf{G}'}(\textbf{q},\omega) = \frac{2}{S}\sum^{SBZ}_{\textbf{k}}\sum_{n}^{occ}\sum_{n'}^{unocc}\frac{f_{\textbf{k},n}-f_{\textbf{k}+\textbf{q},n'}}{E_{\textbf{k},n}-E_{\textbf{k}+\textbf{q},n'}+(\omega+i\eta)}\langle\phi_{\textbf{k},n}|e^{-i(\textbf{q}+\textbf{G})\cdot r}|\phi_{\textbf{k}+\textbf{q},n'}\rangle\langle\phi_{\textbf{k}+\textbf{q},n'}|e^{i(\textbf{q}+\textbf{G}')\cdot r}|\phi_{\textbf{k},n}\rangle,
\end{equation}
\end{widetext}\label{chi0}
where \textit{n} (\textit{n'}) is an occupied (unoccupied) band index, \textbf{k} is in the two-dimensional surface Brillouin zone (SBZ), $f_{\textbf{k},n}$ are Fermi factors and $E_{\textbf{k},n}\ \ (\phi_{\textbf{k},n})$ are Kohn-Sham energies (wave functions). Actually in order to speed up the calculations, following Refs.~\onlinecite{aryaprl94} and \onlinecite{zhukprb01}, first the spectral function is calculated and from its knowledge the imaginary and real parts of $\chi^{0}_{\textbf{G},\textbf{G}'}$ are obtained. Finally, the expression for the surface response function in the case of a periodically repeated slab reads
\begin{equation}
\textsl{g}(\textbf{q},\omega) = - \dfrac{2\pi}{q}\int dz \int dz' \chi_{\textbf{G}=0,\textbf{G}'=0}(z,z',\textbf{q},\omega)e^{q(z+z')},\label{g_per}
\end{equation}
Even though only the $\textbf{G}=\textbf{G}'=0$ matrix element of $\chi_{\textbf{G},\textbf{G}'}$ enters Eq.~(\ref{g_per}), the full three-dimensional (3D) nature of the polarizability is implicity taken into account via the evaluation of Eq.~(\ref{chi_int}) as a matrix equation.

In order to save computational time, $\chi^{0}_{\textbf{G},\textbf{G}'}$ has been calculated retaining only $\textbf{G} = (0,0,G_{z})$ reciprocal space vectors. Physically, this means that lateral crystal local field effects \cite{adlpr62} were neglected. This approach was already found to give indistinguishable results compared with the calculations carried using the 3D \textbf{G'}s for metal surfaces.\cite{silkprl04} All important 3D effects are included in the evaluation of $\chi^{0}_{\textbf{G},\textbf{G}'}$ through the use of the fully 3D Bloch functions and their respective one-electron energies.
 
In the present work Pb(111) films are represented by freestanding slabs infinite in the \textit{xy} plane and periodically repeated in the \textit{z} direction, separated by a vacuum region whose thickness here is fixed in all cases as 10 interlayer distances of the lead atoms of the film in the \textit{z} direction. Films are not relaxed, representing ideal cuts of the face-centered cubic bulk Pb in the (111) direction with the bulk experimental lattice parameter of $4.95\ \mathring{A}$. Thus, the in-plane lattice parameter is $a=3.50\ \mathring{A}$, the interlayer distance is $c=2.86\ \mathring{A}$ and the vacuum region thickness is $d=28.6\ \mathring{A}$. However, 4 - 6 ML thick films were also allowed to relax in the \textit{z} direction and their band structure showed small changes compared with their unrelaxed counterparts.

For the density functional theory (DFT) ground state calculations, the electron-ion interaction is represented by norm-conserving non-local pseudopotentials,\cite{bacheleprb82} and the LDA approximation is chosen for the exchange and correlation potential, with the use of the Perdew-Zunger \cite{pezuprb81} parametrization of the XC energy of Ceperley and Alder.\cite{cealprl80} Well-converged results have been found with a kinetic energy cut-off of $\sim$220 eV, including from $\sim2200$ (1 ML) to $\sim5300$ (15 ML) plane-waves in the expansion of the Bloch states.

For 1 - 4 ML thick films, the Hamiltonian was also solved including the SO term fully self consistently. As a centrosymmetric supercell was used in the calculations, due to the Kramers degeneracy \cite{tinkam71} the electron energy bands are doubly degenerate also when the SO coupling is included in the Hamiltonian (see Fig.~\ref{bs}).

The calculation of $\chi^{0}_{\textbf{G},\textbf{G}'}(\textbf{q},\omega)$ was carried out using a Monkhorst Pack 192$\times$192$\times$1 (96$\times$96$\times$1) grid of {\bf k} vectors as the hexagonal SBZ sampling with 3169 (817) \textbf{k} vectors in the irreducible part of the SBZ for the 1-5 ML (6-15 ML) thick films. Up to 500 bands were included in the evaluation of $\chi^{0}_{\textbf{G},\textbf{G}'}(\textbf{q},\omega)$ for all thicknesses. The width of the Gaussian replacing the delta function in the evaluation of $\chi^{0}_{\textbf{G},\textbf{G}'}(\textbf{q},\omega)$ was set to 0.15 eV, a value which gave smooth results while not hiding any feature on the surface loss function of the films. Well converged results are found including 750 plane waves in the expansion of the wave functions in the calculation of $\chi^{0}_{\textbf{G},\textbf{G}'}(\textbf{q},\omega)$ and expanding the size of the polarizability matrix up to 60 \textbf{G} vectors.

\section{ELECTRONIC STRUCTURE RESULTS}

The results of the ground state calculations showed bilayer oscillations as a function of the slab thickness on the density of states at the Fermi level and on the work function, with a beating pattern of period 9 ML superimposed (not shown). This is in agreement with previous experimental and theoretical studies (see i.e. Refs.~\onlinecite{kimpnas10} and \onlinecite{weiprb02}).

In Fig.~\ref{bs} the calculated electronic band structure of Pb(111) films of several thicknesses is shown. For a N ML thick film, each electron state energy level is unfolded in N subbands. The subbands below -6 eV are of \textit{s} character. They are separated by a gap from the 3N subbands of \textit{p} character which form the Fermi surfaces of the slabs. As can be seen from Fig.~\ref{bs}, the width of the gap is already fixed as $\sim$2 eV for the 3 ML thick film.

Around the SBZ center ($\overline{\Gamma}$ point) bands show a parabolic free-electron-like dispersion. The \textit{p} bands around $\overline{\Gamma}$ present a $p_z$ character, while acquiring an increasing $p_{x,y}$ component as they loss their parabolic-like dispersion moving away from $\overline{\Gamma}$.\cite{kirenp10} The $p_z$ states represent the QWS of the Pb(111) films. The present work found that the inverse of the energy separation of the QWS around $E_F$ is linearly proportional to the film thickness, in good agreement with a previous study.\cite{weiprb02}

As lead is a heavy element (atomic number 82), SO interaction has sizeable effects on its energy spectrum. As an example, in bulk Pb the SO-induced splitting at the BZ center is $\sim$3 eV, and several degeneracies are lifted throughout the BZ.\cite{zubiprb11} In Fig.~\ref{bs} the band structure and DOS for the 1 - 3 ML thick films is shown with (dashed lines) and without (solid lines) SO coupling included in the Hamiltonian. As can be readily seen, SO effects are remarkable only for the single monolayer case, which becomes semimetallic when the SO interaction is switched on, as a result of the avoiding of the band-crossings present for the scalar-relativistic system around the Fermi level. As the slab thickness is increased, also the filling of the phase space by the unfolding of the subbands increases. Because of the fast filling of the phase space, SO effects on the ground state of Pb(111) films become small for slabs as thin as 3 ML (see Fig.~\ref{bs}), as avoiding of the band-crossings is the only sizeable SO effect on their band structure. As a consequence, SO effects are not expected to affect qualitatively the films surface loss function, except for the somewhat artificial semimetallic single Pb(111) monolayer. Thus, in the present work only scalar-relativistic calculations are reported. Note however that a recent first-principles study has shown the inclusion of the SO coupling as necessary in the calculation of the electron-phonon coupling and the superconducting temperature of Pb(111) films.\cite{sklyprb13}

\section{SURFACE LOSS FUNCTION RESULTS}

\begin{figure}[b]
\includegraphics[width=0.48\textwidth]{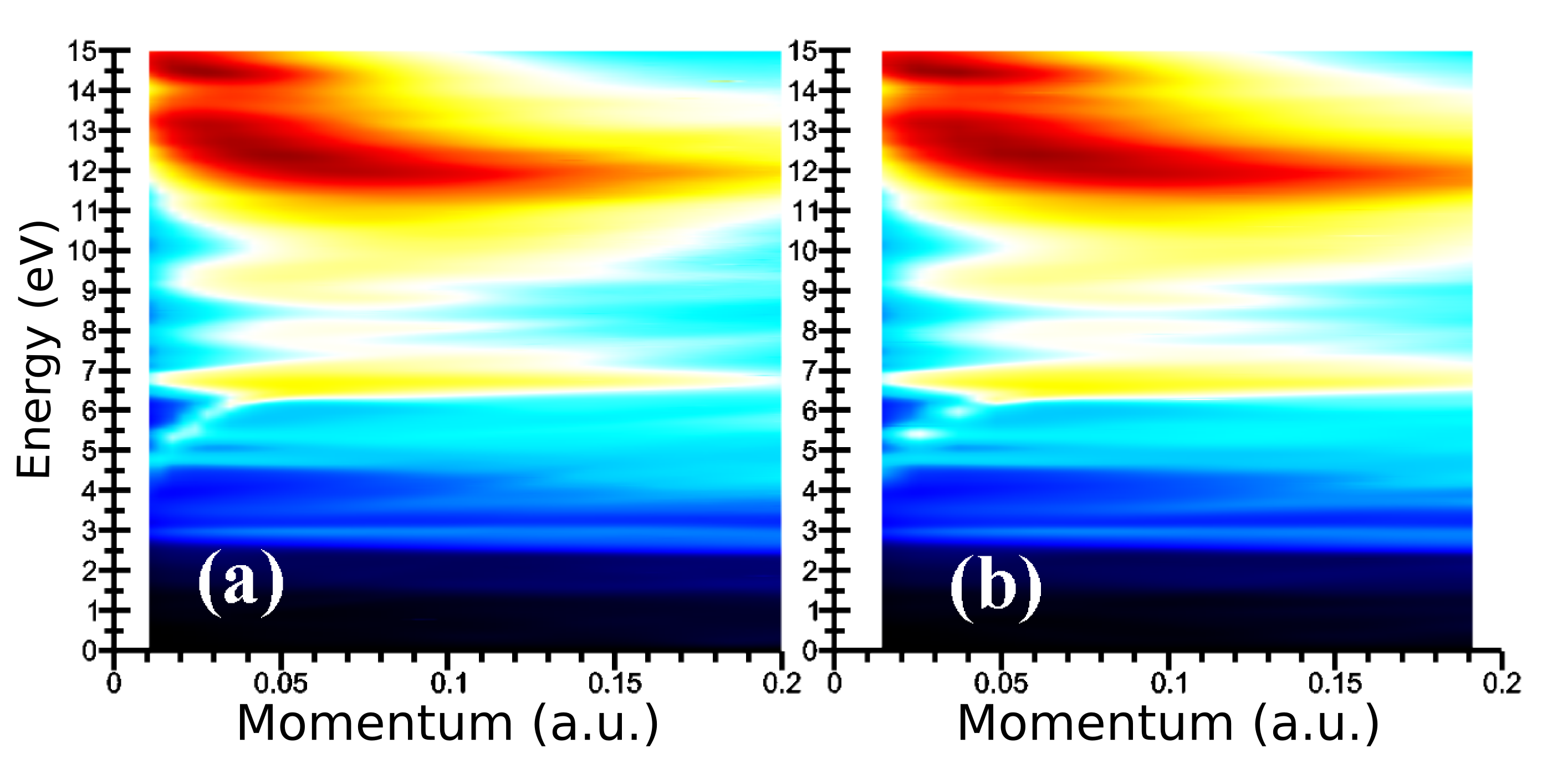}
\caption{(Color online) Calculated surface loss function for the 3MLs thick Pb(111) film, with $\textbf{q}$ along two different high-symmetry directions, $\overline{\Gamma}-\overline{M}$ (a) and $\overline{\Gamma}-\overline{K}$ (b)}.\label{ani}
\end{figure} 

\subsection{Isotropy}

\begin{figure*}[t]
\includegraphics[width=1.0\textwidth]{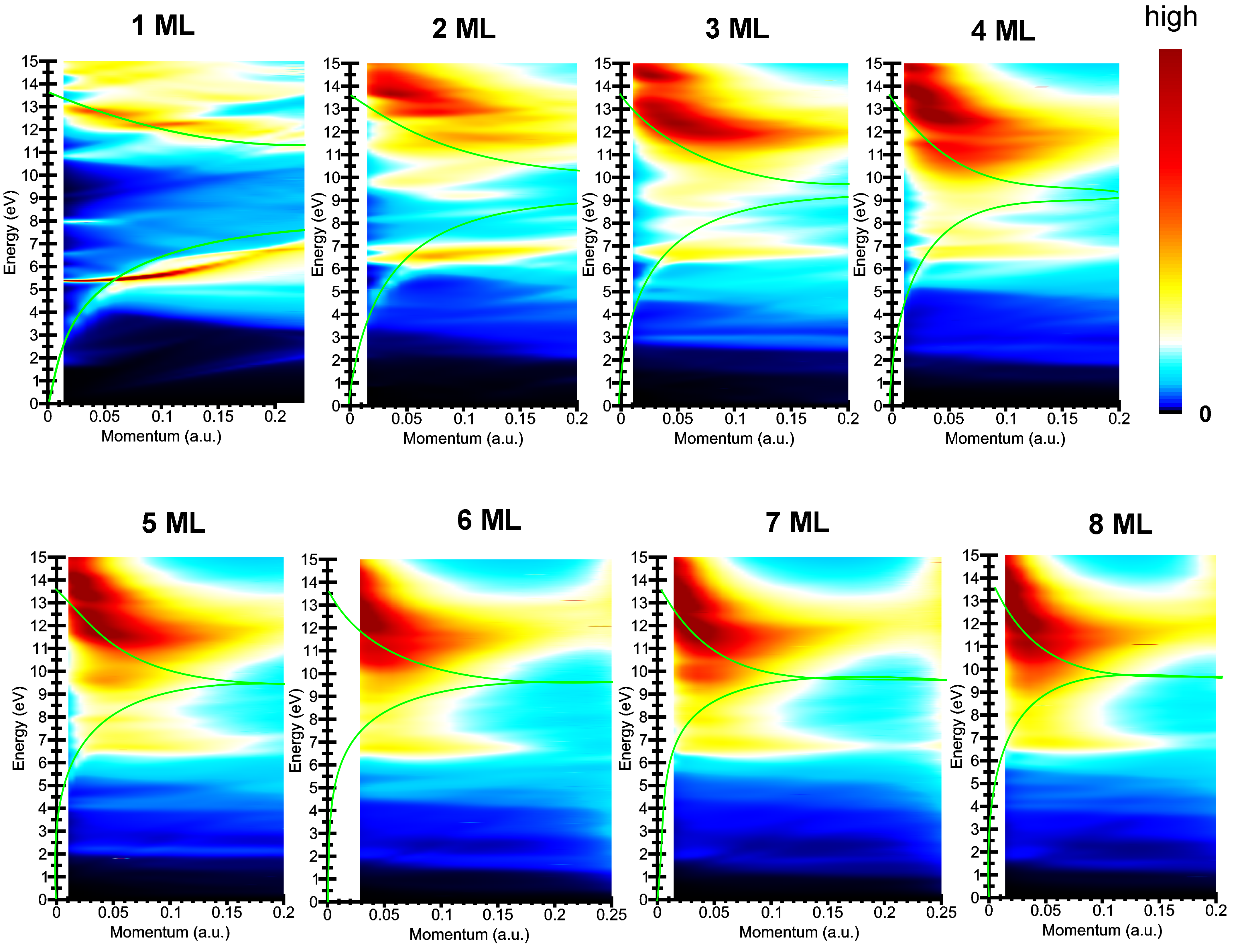}
\caption{(Color online) Surface loss function surfaces Im[$\textsl{g}(\textbf{q},\omega$)] for 1 - 8 MLs thick Pb(111) films calculated at the scalar-relativistic level. The momentum transfer vector $\textbf{q}$ is along $\overline{\Gamma}-\overline{M}$. The colour bar applies to all plots, with its maximum normalized individually for each case. Solid lines represent the dispersion of the classical modes, see Eq.~(\ref{classical_dispersion}).} \label{main}
\end{figure*}

In Fig.~\ref{ani} the calculated surface loss function for the 3MLs thick Pb(111) film, with $\textbf{q}$ along two different high-symmetry directions, namely $\overline{\Gamma}-\overline{M}$ [panel (a)] and $\overline{\Gamma}-\overline{K}$ [panel (b)] is shown. It is clear that Im[$\textsl{g}(\textbf{q},\omega)$] exhibits a highly isotropic character. In all the carried tests the same isotropic behaviour of the surface loss function was found independently of the film thickness. Thus, from here on only results for $\textbf{q}$ along $\overline{\Gamma}-\overline{M}$ are shown, as the grid used in this high-symmetry direction is finer than the one along $\overline{\Gamma}-\overline{K}$.

\subsection{General results}

The general results of the present work are shown in Fig.~\ref{main}. In order to get insight, the dispersion of the classical modes $\omega_{\pm}=\omega_{\pm}(q)$ given by Eq.~(\ref{classical_dispersion}) is represented by solid green curves. For each freestanding slab, $\omega_{\pm}(q)$ are plotted for an effective thickness corresponding to a number of interlayer distances equal to the number of MLs forming the slab, as the jellium edge in the first-principles calculations was fixed at half an interlayer distance away from the outermost atomic layers.

The results of the present work as plotted in Fig.~\ref{main} show several modes of different character. First, the low-energy symmetric mode is detected for the thinnest slabs at small momentum transfer values, closely following the dispersion described by the low-energy $\omega_{-}$ mode of Eq.~(\ref{classical_dispersion}) for all thicknesses as represented in Fig.~\ref{main} by the bottom green line in each panel. However, notice that it disappears upon entering the almost dispersionless peak present around $\omega \simeq 7$ eV for all thicknesses.

Also, the high-energy plasmon mode analogous to the classical thin film $\omega_{+}$ mode is found for thicknesses greater than 2MLs. Note that it is placed at too high energies in comparison with the predictions of Eq.~(\ref{classical_dispersion}). Unfortunately, calculations including the 5$d$ semicore electrons are too computationally demanding in the supercell scheme used here. Thus, the influence of the semicore electrons on Im[$\textsl{g}(\textbf{q},\omega$)] has not been checked . The inclusion of the polarizable 5$d$ semicore electrons through the use of a model dielectric function $\varepsilon_d$ \cite{liebsch97} is ambiguous and its use has been discarded in this present study. Note the 5$d$ electrons have been shown to play a crucial role in the dielectric response of bulk Pb, more precisely in the optics and dynamics of the main bulk plasmon.\cite{zubiprb13b}

\begin{figure*}[t]
\includegraphics[width=0.9\textwidth]{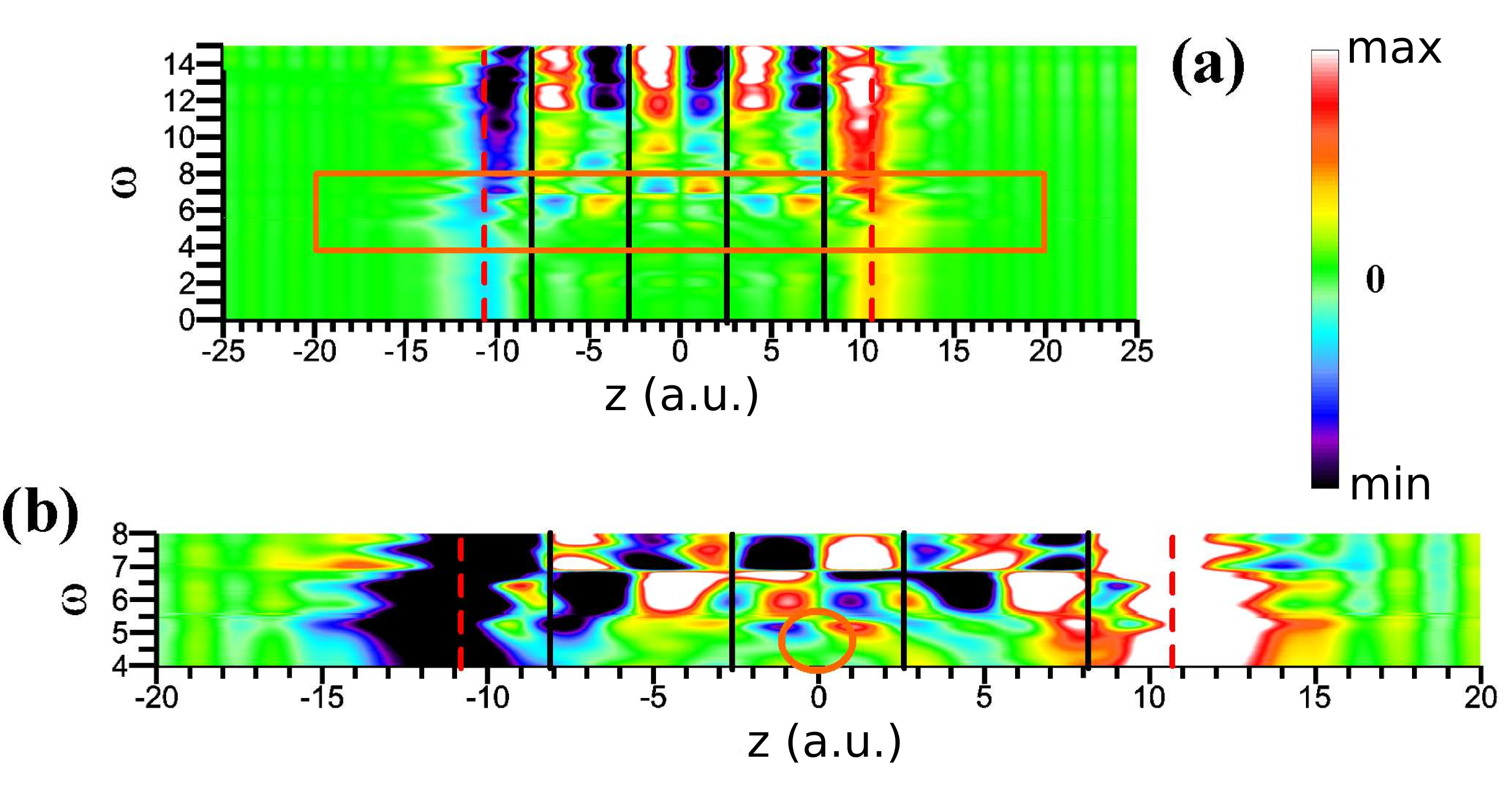}
\caption{(Color online) Real part of the induced density at $q$=0.014 a.u. for the 4 MLs Pb(111) film, as a function of the energy transfer $\omega$ and the $z$ coordinate. Panel (b) is a zoom in the ($\omega,z$) range delimited by the orange rectangle in panel (a). The orange circle in (b) highlights the fingerprint of the low-energy symmetric mode (see the text). Black solid (red dashed) lines mark the position of the atomic layers (jellium edges). $\omega$ in eV.} \label{n_ind}
\end{figure*} 

\begin{figure}[b]
\includegraphics[width=0.45\textwidth]{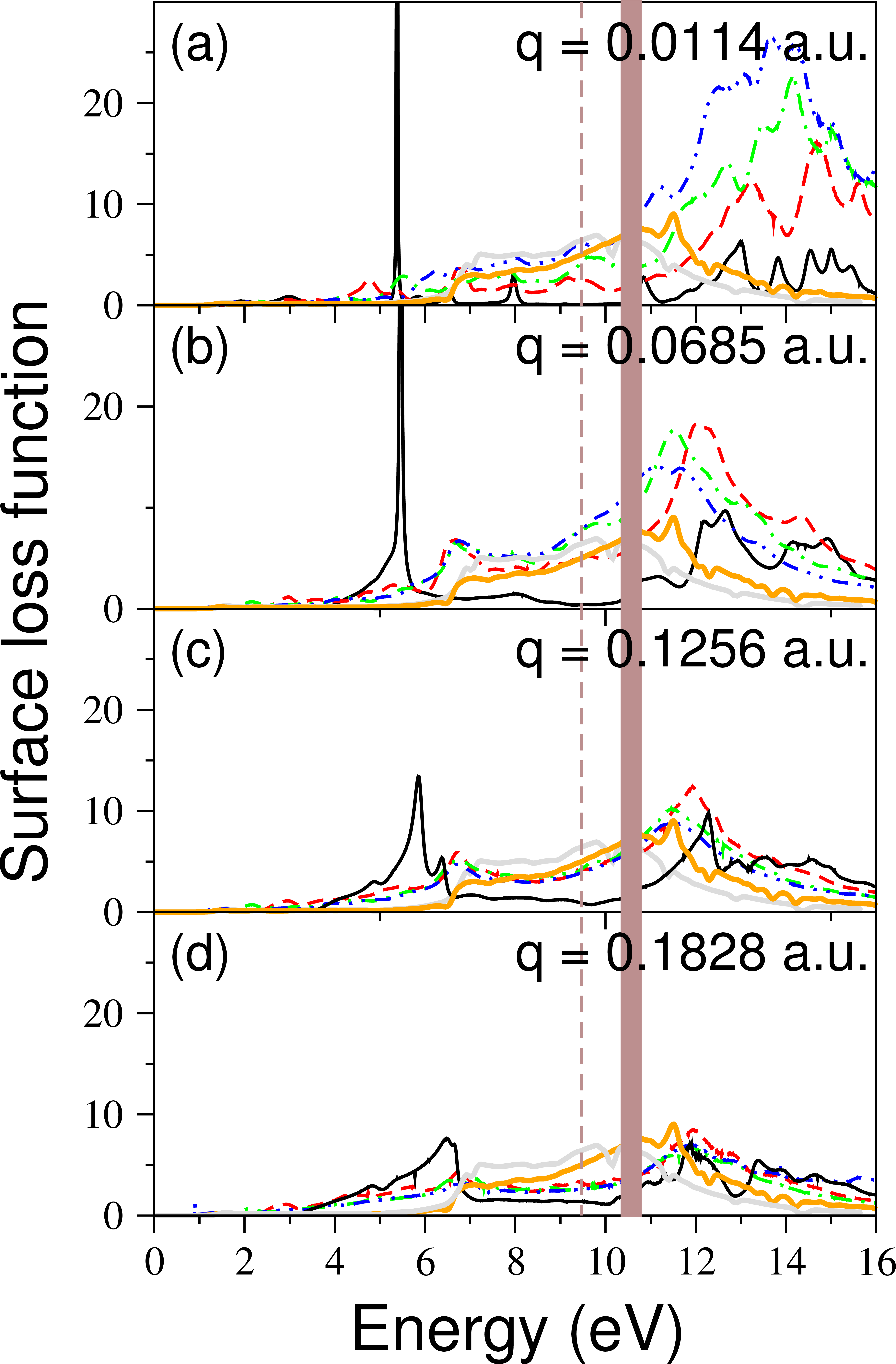}
\caption{(Color online) Surface loss function for different values of $\textbf{q}$ along $\overline{\Gamma}-\overline{M}$ and different film thicknesses. Black solid, red dashed, green dashed-dotted and blue dashed-dotted-dotted curves represent results for 1, 3, 5, and 8 MLs thick Pb(111) films, respectively. The thick orange (grey) solid curve stands for the results deduced from bulk calculations without (with) inclusion of the 5$d$ electrons. The vertical dashed line marks the classical Pb surface plasmon energy of 9.57 eV, while the shaded energy interval corresponds to the electron energy loss experimental value of 10.6$\pm$0.2 eV.\cite{powepps60}}\label{thick}
\end{figure}

In Ref.~\onlinecite{yuanprb06} using jellium calculations it was shown that the antisymmetric mode disappears for film thicknesses comparable or smaller than the Fermi wavelength of the metal when $q \rightarrow 0$. Instead, peaks corresponding to discrete interband transitions show up. For Pb, using the value $r_s^{\rm Pb}=2.298$, one finds $\lambda_F^{\rm Pb} = 7.52$ a.u. which is roughly 1.4 times the interlayer distance in Pb(111) films. Thus, the 1 and 2 MLs thick Pb(111) films present electronic effective thicknesses equal to 0.7 and 1.4 times $\lambda_F^{\rm Pb}$, respectively. As seen in Fig.~\ref{main}, our results are in agreement with the work of Ref.~\onlinecite{yuanprb06} as far as the disappereance of the high-energy mode for thin films is concerned. In the surface loss function of the single monolayer shown in Fig.~\ref{main}, a manifold of interband peaks is present for energy transfers $\omega \gtrsim 11$ eV [see also the black solid curve in Fig.~\ref{thick} (a)], where the high-energy mode should be present (see the upper green line in the first panel of Fig.~\ref{main}). This is a manifestation of strong QSE in the surface-loss function of the single Pb(111) monolayer. The 2 ML Im[$\textsl{g}(\textbf{q},\omega$)] results (see Fig.~\ref{main}) correspond to the transition between the two different thickness regimes, at $L\simeq1.4\lambda_F^{\rm Pb}$.

An important conclusion of the present work is the large difference in spectral weight between the low- and high-energy modes of the film, in sharp contrast with the results reported in Ref.~\onlinecite{yuanprb06} for Ag slabs modeled by the jellium approximation. The low-energy mode analogous to the classical symmetric $\omega_{-}$ plasmon appears as a faint feature in comparison with the rest of the peaks present in Im[$\textsl{g}(\textbf{q},\omega$)]. On the contrary, the high-energy mode is the most intense feature in the surface loss function of freestanding Pb(111) films, except for the single monolayer. In the latter case, a slightly upwards dispersing interband peak raises at energies $\omega \simeq 5.5 - 6$ eV. It exhibits the highest intensity [see Fig.~\ref{thick} (a) and (b)] together with a vanishing linewidth at momentum transfer smaller than 0.1 a.u. This long-living mode stems from transitions between the highest occupied and lowest unoccupied QWSs around the SBZ center (see Fig.~\ref{bs}), representing strong QSE. Once more the 2 MLs results represent the crossover with larger thicknesses for which the quantization of the states is not reflected in the same fashion in the calculated surface loss function. Nevertheless, in the evaluated Im[$\textsl{g}(\textbf{q},\omega$)] corresponding to the 2 MLs thick slab, still two interband peaks similar to the long-living mode present in the single monolayer are found, overlapping with each other. However, their intensity is greatly decreased in comparison with the corresponding feature of the response of the 1ML Pb(111) film.

In Fig.~\ref{n_ind} the real part of the two-dimensional Fourier transform of the induced density [see Eq.~(\ref{nind})] Re[$n^{\rm{ind}}(z,\textbf{q},\omega)$] for the 4 MLs thick slab is shown. The results correspond to a momentum transfer of $q$ = 0.014 a.u. First, note the antisymmetric distribution of the induced density with respect to $z$=0. Second, for spatial positions inside the film, several changes of the sign of Re[$n^{\rm{ind}}$] are found. In panel (b) results for $4\leqslant\omega\leqslant8$ eV are zoomed in. Interestingly a sharp change of phase of Re[$n^{\rm{ind}}$] can be recognized at $\omega \simeq 7$ eV, signaling about the presence of the dispersionless peak seen in Im[$\textsl{g}(\textbf{q},\omega$)] at this energy. This is a general behaviour found at all thicknesses at $\omega \simeq 7$ eV.

Moreover, the fingerprint of the low-energy mode analogous to the classical symmetric $\omega_{-}$ plasmon is found, as marked by the circle in panel (b) of Fig.~\ref{n_ind}. As can be seen, $\omega \sim$ 5 eV is the only energy at which there is a noticeable weight of Re[$n^{\rm{ind}}(z,\textbf{q},\omega)$] at the center of the slab, $z$ = 0. This notable distortion in the general antisymmetric distribution of the real part of the two-dimensional Fourier transform of the induced density signals about the presence of the symmetric plasmon mode. This faint but appreciable $\omega_{-}-$like fingerprint has been found at all thicknesses in which the symmetric mode could be resolved.

\subsection{Thickness dependence}

In order to relate the results for different thicknesses, several cuts of Im[$\textsl{g}(\textbf{q}_\parallel,\omega$)] are plotted in Fig.~\ref{thick} comparing the surface loss function of Pb(111) films of distinct thicknesses for the same momentum transfer values of $\textbf{q}$ along the $\overline{\Gamma}-\overline{M}$ high-symmetry direction.

In panel (a) of Fig.~\ref{thick} the black curve at $\omega \gtrsim 11$ eV shows the manifold of interband peaks which replaces a single high-energy antisymmetric mode for the Pb monolayer. An additional important feature in the surface loss function results for the 1 ML slab is the long-living interband peak found at small momentum transfer, seen at $\omega = 5.5$ eV in panels (a) and (b) of Fig.~\ref{thick}.

For the 3, 5 and 8 MLs thick films the surface plasmon is already present at $q$=0.1256 a.u., as seen in panel (c) of Fig.~\ref{thick}. Note that it presents a remarkably smaller intensity than the high-energy antisymmetric mode. On the other hand, the low-energy symmetric mode can not be seen in the scale of Fig.~\ref{thick}, as it is a faint feature (see Fig.~\ref{main}).

\subsection{Surface plasmon}

\begin{figure}[b]
\includegraphics[width=0.45\textwidth]{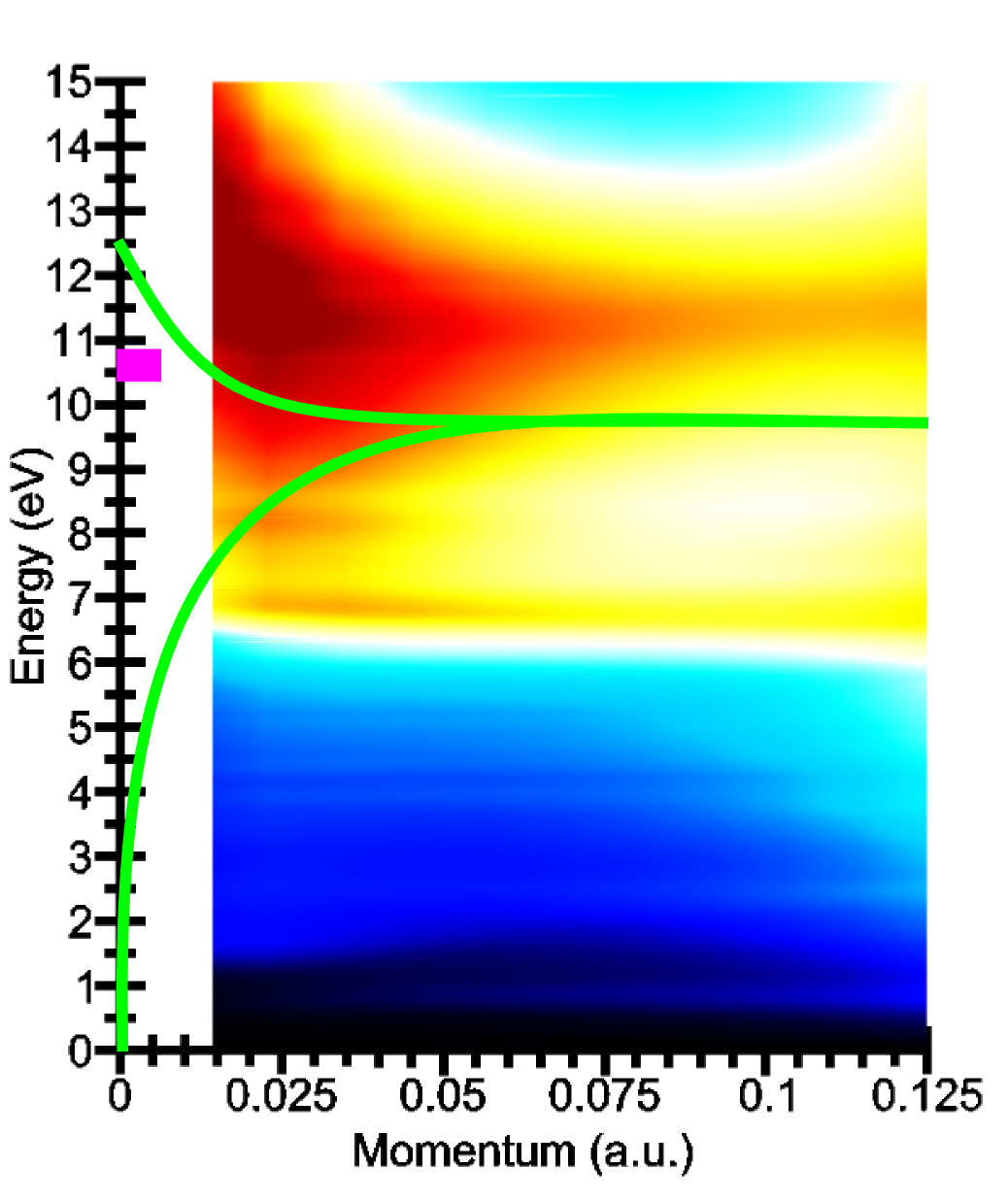}
\caption{(Color online) Scalar-relativistic surface loss function of the 15 MLs thick Pb(111) film, $\textbf{q}$ along $\overline{\Gamma}-\overline{M}$. The green lines stand for the dispersion of the classical thin film modes as given by Eq.~(\ref{classical_dispersion}). The pink square represents the experimental data of Ref.~\onlinecite{powepps60} of $\omega_{s}^{\rm{exp}}=10.6\pm0.2$ eV. The same colour code as in Fig.~\ref{main} is used.} \label{15ML}
\end{figure}

In Fig.~\ref{thick} the vertical dashed line marks the classical surface plasmon energy $\omega_{s}=\omega_{p}/\sqrt{2}=\sqrt{1.5r_s^{-3}}$, which for the averaged valence electron density of bulk lead $r_{s}^{\rm Pb}=2.298$ gives the value $\omega_{s}^{\rm Pb}$ = 9.57 eV. Also, the results of the experimental electron energy-loss measurements of 10.6$\pm$0.2 eV \cite{powepps60} are represented by the thin shaded area. As can be seen, the classical expression gives a too low value of the surface plasmon energy by about 1 eV.

On the other hand, in the optical limit ($q \rightarrow 0$) the surface response function can be calculated from the bulk dielectric function as \cite{liebphs87,liebsch97}
\begin{equation}
\textsl{g}(q\rightarrow0,\omega) = \frac{\varepsilon^{\rm{bulk}}(q\rightarrow0,\omega)-1}{\varepsilon^{\rm{bulk}}(q\rightarrow0,\omega)+1},\label{g_opt}
\end{equation}
and thus the surface loss function is
\begin{equation}
\rm{Im}[\textsl{g}(q\rightarrow0,\omega)] \propto - \rm{Im}\left[\frac{1}{\varepsilon^{bulk}(q\rightarrow0,\omega)+1}\right].\label{im_g_opt}
\end{equation}

In Fig.~\ref{thick} the orange (grey) thick solid curve represents $\rm{Im}[\textsl{g}(q\rightarrow0,\omega)]$ calculated using Eq.~(\ref{im_g_opt}) and including the 5$d$ electrons in the core (valence). The energy of this peak at half width at half maximum (HWHM) is of 10.85 (9.3) eV with the semicore electrons excluded from (included in) the valence configuration. When using the bulk dielectric function obtained without including the 5$d$ electrons, the value retrieved is close to the experimental one of 10.6$\pm$0.2 eV. However, the agreement is worsen upon taking the semicore electrons into account in the evaluation of $\varepsilon^{bulk}(q\rightarrow0,\omega)$. Note that the surface loss function obtained from a bulk calculation (without the semicore) through Eq.~(\ref{im_g_opt}) is in qualitative agreeement with the slab surface plasmon for thicknesses greater than 2 MLs at momentum transfer values where the modes $\omega_{\pm}$ are uncoupled, see panel (c) in Fig.~\ref{thick}.
\begin{figure}
\includegraphics[width=0.48\textwidth]{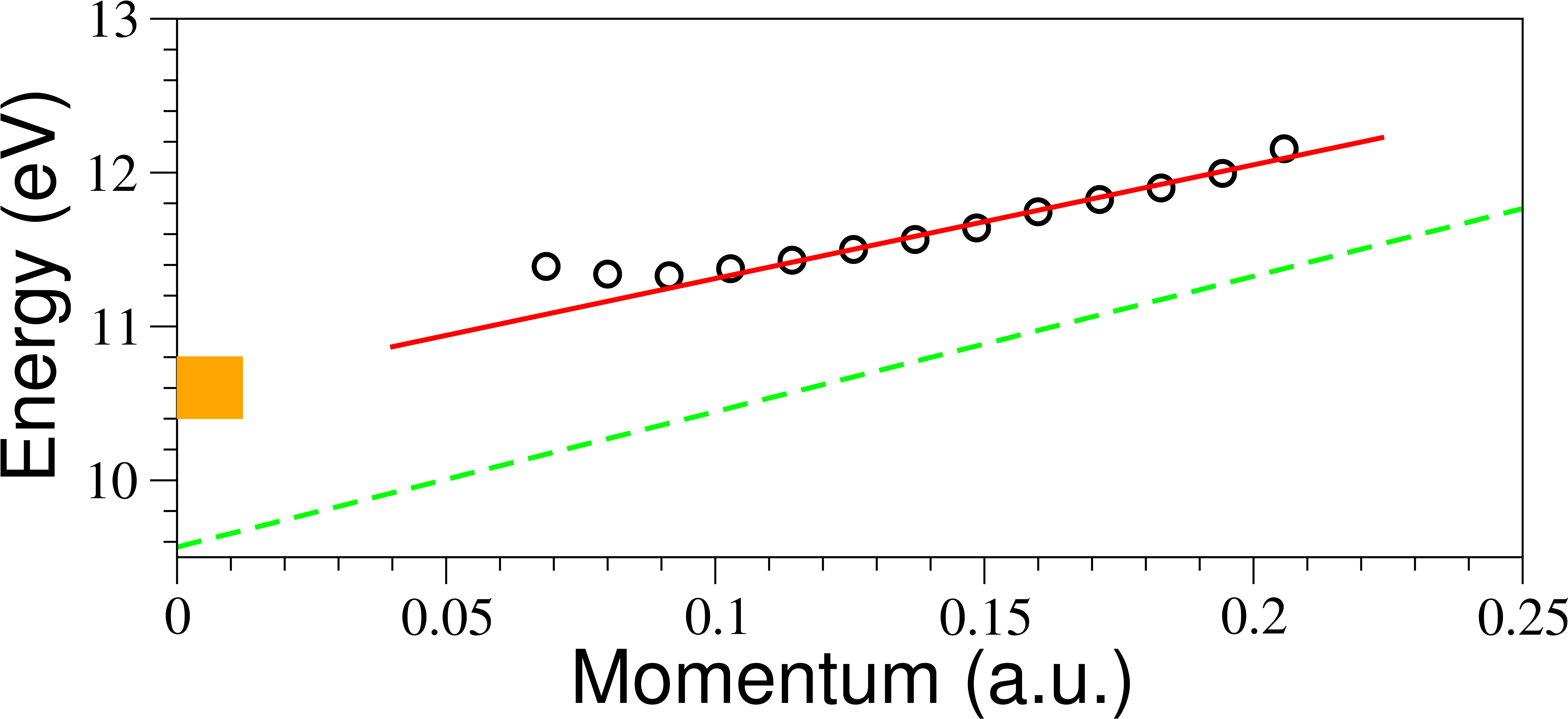}
\caption{(Color online) Surface plasmon dispersion for the 7 MLs thick Pb(111) film as a function of $\textbf{q}$ (along $\overline{\Gamma}-\overline{M}$). The circles represent the calculated values of the HWHM position at each \textit{q}. The red solid line is a linear fit of the computational results, while the green dashed one stands for the surface plasmon dispersion in a hydrodynamic approach of the jellium semiinfinite surface (see the text). The shaded orange square marks the experimental interval of $\omega_{s}^{\rm{exp}}=10.6\pm0.2$ eV.\cite{powepps60}} \label{7ML}
\end{figure} 
Surprisingly, Im$[\textsl{g}(q\rightarrow0,\omega)]$ calculated from the knowledge of $\varepsilon^{bulk}(q\rightarrow0,\omega)$ shows a faint peak at 7 eV, mimicking the dispersionless feature which plays the role of the short wavelength limit of the symmetric mode $\omega_{-}$ in the thinnest films. This signals about the bulk-like character of the aforementioned dispersionless interband mode.

From Figs.~\ref{main} and \ref{15ML}, it seems the surface plasmon disperses roughly linearly with the momentum transfer. In Fig.~\ref{7ML} the calculated dispersion $\omega_{s} = \omega_{s}(q)$, with $\textbf{q}$ along $\overline{\Gamma}-\overline{M}$, is shown for the 7 ML thick Pb(111) film. Thicker films did not present any remarkable difference in the surface plasmon dispersion. The values of $\omega_{s}$ were evaluated at the position of the HWHM. As can be seen, the surface plasmon presents a fairly linear dispersion as a function of the momentum transfer for $q \gtrsim 0.1$ a.u.

The straight line in Fig.~\ref{7ML} is the result of fitting $\omega(q) = A + B\cdot q$ for $q > 0.1$ a.u. The obtained values of the fitting parameters are $A$ = 10.58 eV and $B$ = 7.35 eV$\cdot$a.u. It is interesting to compare this findings with a simple model giving a similar behaviour of $\omega_s = \omega_s(q)$.

In a semiinfinite jellium surface, using the so-called on-step hydrodynamic approach,\cite{pitarpp07,lundthe83} the following expression for the dispersion of the surface plasmon is found at long wavelengths:
\begin{equation}
\omega_{s}(q) = \frac{\omega_{p}}{\sqrt{2}} + \frac{\beta q}{2}, \label{hydro}
\end{equation}
where $\omega_{p}/\sqrt{2}=\sqrt{1.5r_{s}^{-3}}$ and $\beta=\sqrt{3/5}(v_{F}/2)$ \cite{pitarpp07,lundthe83}, being $v_{F} = (9\pi/4)^{1/3}r_{s}^{-1}$ the Fermi velocity of a free-electron gas of average valence electron density parameter $r_s$. Using $r_{s}^{\rm Pb}$ one gets $\omega_{p}^{\rm Pb}/\sqrt{2} = 9.57$ eV and $\beta^{\rm Pb}/2$ = 8.801 eV$\cdot$a.u. This dispersion is plotted in Fig.~\ref{7ML} as a green dashed line, while the orange square shows the energy interval for the experimentally determined value of $\omega_{s}^{\rm exp}=10.6\pm0.2$ eV.\cite{powepps60} The dispersion derived from the hydrodynamic approach fails in reproducing a correct value for the optical surface plasmon energy (as pointed above). Note that strictly speaking, Eq.~(\ref{hydro}) is valid for $q\ll2\omega_{p}/\beta$.\cite{pitarpp07,lundthe83} In the case of lead, this gives the condition $q \ll 1.087$ a.u.

Finally, note it is difficult to deduce a value of $\omega_s(q\rightarrow0)$ from the present calculations, as the surface plasmon disperses with the momentum transfer in contrast to the classical picture described by Eq.~(\ref{classical_dispersion}). In addition, the well-known negative dispersion of the surface plasmon as a function of $\textbf{q}$ in the long wavelength limit is not retrieved in the present work, as even for the thickest film studied (15 MLs) the low- and high- energy modes are splitted for the smallest values of $\textbf{q}$ used.

\section{CONCLUSIONS}

In the present work, the surface loss function of thin Pb(111) films has been studied by means of a first-principles pseudopotential approach using a supercell scheme.

For 1 and 2 MLs thick films strong QSE have been found. The high-energy mode is completely absent in the dielectric response of the single monolayer. This is a direct consequence of the quantization of the electronic states, leading instead to the appearance of discrete interband transitions in the high-energy range at small momentum transfer \textit{q} [see Fig.~\ref{thick} (a)]. This is in agreement with a previous work based on the jellium model.\cite{yuanprb06}

Incorporation of the full 3D \textit{ab initio} band structure also shows a new feature. It does not disperse with the momentum transfer for films thicker than 2 MLs, presenting an energy of $\omega \sim 7$ eV. In practice, this new mode plays the role of the classical surface plasmon as the long-$q$ limit of the low-energy thin film mode, as $\omega_{-}$ disappears upon coupling to the dispersionless peak. To the best of our knowledge, this is the first work predicting the existence of this new mode as the short wavelength limit of the low-energy mode, replacing the role of the classical surface plasmon of energy $\omega_{s}=\omega_{p}/\sqrt{2}$. Indeed, in Ref.~\onlinecite{powepps60} a value of 7.2$\pm$0.1 eV was reported as the average energy of a feature below the surface plasmon energy in EELS measurements. We identify this feature as the dispersionless mode found in the present \textit{ab initio} study. Surprisingly, the optical surface loss function evaluated from bulk calculations [see Eq.~(\ref{g_opt})] also shows a faint peak at $\sim$ 7 eV.

Also, the surface loss function calculated from Eq.~(\ref{g_opt}) is in agreeement with the first-principle results [see Fig.~\ref{thick} (c) and (d)]. As regards the surface plasmon dispersion, a linear dependence with $q$ has been found in the present work. Once its dispersion is fitted to a linear function of the momentum transfer, extrapolation of the fitting to $q \rightarrow 0$ gives a value of 10.58 eV, in agreement with the experimental $\omega_{s}^{\rm exp}$ = 10.6$\pm$0.2 eV.\cite{powepps60}

New electron energy loss spectroscopy measurements on Pb(111) thin films are highly desirable to check the present predictions and gain further insight in the dynamics of collective electronic excitations of nanostructured systems and the consequences of the quantization of the electronic states on them.

\section*{ACKNOWLEDGEMENTS}

We are grateful to I\~{n}igo Aldazabal for technical help in computational optimization. We also acknowledge financial support from the Spanish MICINN (No. FIS2010-19609-C02-01), the Departamento de Educaci\'on del Gobierno Vasco, and the University of the Basque Country (No. GIC07-IT-366-07).

\end{document}